\documentclass[twocolumn]{aa}
\usepackage{txfonts}
\usepackage{graphicx}
\usepackage{natbib}
\def\aj{{AJ}}                   
\def\araa{{ARA\&A}}             
\def\apj{{ApJ}}                 
\def\apjl{{ApJ}}                
\def\aap{{A\&A}}                
\def\mnras{{MNRAS}}             

\def\ms{{\mathcal S}}

\begin{document}

\title{Colour dependence of the distribution of IRAS galaxies revealed by multifractal measures}

\author{Jun Pan \thanks{\emph{e-mail address:}Jun.Pan@nottingham.ac.uk}}

\institute{Purple Mountain Observatory, Chinese Academy of Sciences, Nanjing 210008, China
\and
School of Physics \& Astronomy, Nottingham University, Nottingham NG7 2RD, UK
}

\date{Received/ Accepted}
\abstract{
Multifractal measures are applied to three IRAS galaxy subsamples selected by colour
from the PSCz catalogue. As shown by a generalised dimension spectrum, hot IRAS galaxies are found to be
less clustered than cold galaxies, but the difference is very small. An alternative tool, the conditional 
multifractal dimension spectrum reveals an apparent peculiarity of the
distribution of hot galaxies, especially at high orders. The conditional multifractal measure 
basically measures the environment of selected galaxies. A detailed analysis of the distribution of
galaxies with their number of neighbours shows that cold galaxies
are more likely to be in over-dense regions than hot galaxies. Further studies suggest that
since the colour of IRAS galaxies is a good trace of star formation rate, 
it is possible that we have statistical evidence here for enhanced star formation rate 
due to galaxy interactions.
\keywords{Cosmology: large-scale structure of Universe -- Infrared: galaxies --
Methods: statistical}
}

\titlerunning{Distribution of IRAS galaxies revealed by multifractal measures}
\maketitle


\section{Introduction}

Galaxies are inevitably affected by their environment. It has been found that 
early-type galaxies are far more common than spirals in centres of rich clusters, 
while the opposite is true in other parts of the universe 
\citep{Dressler1980, CaonEinasto1995, DresslerEtal1997}. 
There is the same phenomenon in galaxies' spatial distribution: 
early-type galaxies are more strongly clustered
than late-types \citep[e.g.,][]{HermitEtal1996, ShepherdEtal2001, 
MadgwickEtal2003}, and high luminosity galaxies are more strongly 
clustered than low luminosity ones \citep[e.g.,][]{HermitEtal1996,
LinEtal1996, GuzzoEtal2000, NorbergEtal2001}.
It is clear that galaxies are biased tracers of the mass, which has invoked the 
concept of bias to make connections between statistics of galaxies and 
the mass \citep[e.g.,][]{Kaiser1984, BardeenEtal1986, MoWhite1996, DekelLahav1999}.
Galaxy samples selected by different criteria are frequently studied to quantify the
phenomenon \citep[e.g.,][]{NorbergEtal2002a, ZehaviEtal2005}.

Among the many galaxy surveys, the IRAS galaxy sample plays an important role in 
studying cosmic large-scale structure and galaxies' formation and evolution. 
Observationally, it affords large sky 
coverage, uniform flux calibration, good position accuracy, and 
insignificant galactic absorption. IRAS galaxies usually are good 
objects of strong star formation activities. The majority of galaxies observed 
in the infrared band are spirals, which tend to avoid rich galaxy clusters, so in general 
the infrared selected galaxy samples have a lower clustering amplitude.

The infrared emission from galaxies are produced by their gas (dust)
component, illuminated mainly by UV emission from stars. 
\citet{RowanCrawford1989} decomposed the spectrum of IRAS galaxies 
into three components: (1) a cool disk component emission from interstellar 
dust activated by galaxy starlight; (2) a warmer ``starburst'' component
interpreted as a burst of star formation in the galaxy
nucleus, whose spectrum is well fitted by a model of
hot stars embedded in optically thick dust cloud; and (3) a ``Seyfert'' 
component that originates in a power-law continuum source within
a dust cloud related to the narrow-line region of the compact source. 
The peak of the disk component's radiation is at the $100\mu$m and the ``starburst''
part reaches maximum at $60\mu$m, while the ``Seyfert'' component's peaks are
at $12\mu$m and $25\mu$m; thus, the far-infrared radiation from the IRAS galaxies 
is a composition of ambient cirrus emission and localised emission from 
regions of active star formation. The relative contribution 
of the two components may be quantified by the dust temperature
inferred from the flux ratio $S_{100}/S_{60}$.

Therefore, $S_{100}/S_{60}$, or equivalently the dust temperature, is a
good measure of the star formation rate (SFR) of IRAS galaxies 
\citep{SaundersEtal1990, MannSaundersTaylor1996}.
There are arguments that in deep gravity potential wells, SFRs of galaxies
are effectively suppressed \citep[e.g.,][]{YoungEtal1996, BaloghEtal1998, 
BlantonEtal1999}; naturally, the segregation phenomenon in clustering is 
expected in the temperature or the SFR selected IRAS galaxy 
subsamples, which is supported by the observation that late-type optical 
galaxies with higher SFRs have smaller clustering amplitudes than early-type 
galaxies. 

Surprisingly, \citet{MannSaundersTaylor1996} found an unusual clustering scenario of 
their ``warm'' and ``cold'' subsamples of QDOT divided by the
$36$K temperature criteria. Their ``warm'' galaxies are more strongly clustered
than the ``cool'' galaxies, which is opposite to what one would expect. However, 
\citet{HawkinsEtal2001} split the PSCz catalogue, which is superior to QDOT, 
into three subsamples, ``hot'', ``warm'', and ``cold'', which fall into temperature ranges 
$>31$K, 28K$\sim 31$K, and $<28$K, respectively, and they found that the hot 
galaxies are actually clustered less strongly than those cold galaxies on scales between
$1h^{-1}$Mpc and $10h^{-1}$Mpc, though the trend is weak.

Here, the problem of the colour dependence of the IRAS galaxies' clustering
is revisited using a multifractal tool. For many years, fractal dimensions were
used to test if the distribution of galaxies in the universe is homogeneous.
Actually, it could have more applications in modern statistical cosmology
than simply as a challenge to the Cosmological Principle.
The benefit of using multifractal is that we have a whole
spectrum of dimensions to describe the distribution. Actually,
in the past, the fractal was already used to study morphology segregation
\citep{WenDengXia1989, DominguezMartinez1989, DominguezEtal1994, BestEtal1996}. 
In addition to conventional fractal dimensions, a new tool, conditional
dimension, was designed to measure differences between two samples. The idea of
conditional dimensions was inspired by concepts of conditional
and relative multifractal spectra \citep{RiediScheuring1997}. Details of the
subsamples' construction are given in Sec. 2. In Sec. 3 we present results from 
ordinary multifractal measurement; Sec. 4 is dedicated to new statistics, the conditional 
multifractal measure. Conclusions and discussions are in the last section.

\section{Sample construction}
\subsection{The main sample from PSCz}
The PSCz catalogue contains 15411 galaxies covering $84\%$
of the sky with $60\mu$m flux, $S_{60}>0.60$ \citep{SaundersEtal2000};
redshifts are available for 14677 galaxies
with a redshift median of $\sim 0.028$. Those galaxies with 
galactic latitude $b<10^\circ$ are thrown away to exclude the high galactic 
absorption plane. Further selection criteria are applied by the galaxy's co-moving 
distance calculated with $H_0=100h^{-1}{\rm kms}^{-1}$ and $\Omega=1$, which is 
$10h^{-1}{\rm Mpc}<r<250h^{-1}{\rm Mpc}$ to avoid the Galaxy's local influence
and make the sample in the remote end not too sparse. 
The final main sample for this work has 11463 galaxies.

We find that the colour of galaxies shows an apparent radial gradient.
It is known that IRAS galaxies experience rapid luminosity evolution with redshift;
we need to check if the colour radial gradient is an evolution effect.
At $60\mu$m, \citet{SaundersEtal1990} gives $L_{60}\propto (1+z)^{3\pm 1}$. Although 
the exact luminosity evolution at $100\mu$m is not clear, we can use 
the luminosity evolution at $90\mu$m, which is $L_{90}\propto (1+z)^{3.4\pm 1}$ 
\citep{SerjeantEtal2004}, as an approximation, so the colour is roughly proportional to
$(1+z)^{0.4}$. The largest redshift
of the sample is $0.089$ which introduces correction to the colour by a factor
of $\sim 1.035$ for the farthest galaxies. Even by this factor, 
since the mean colour of those remote galaxies is $\sim 1.1$ of our sample, 
the modification to the colour is about $0.04$, which 
is relatively small. In fact, after applying such correction to
the colour, changes to the straight line fitting the 
$\log \langle S_{100}/S_{60} \rangle$--$\log r$ relation
are negligible. The colour undergoes very weak evolution in the redshift regime 
of the sample in analysis here. 

Because the main sample is flux-limited, those faint galaxies do not
enter the list, and the mean luminosity will obviously increase with distance.
Actually, there is a correlation between luminosity and
colour: less luminous galaxies are likely to have higher 
colours. Therefore, those galaxies with high colour are unlikely to be selected in
our main sample. The mean colour is low in places at a large distance, while
the radial gradient of colour of the main sample is a selection effect.
\citet{MannSaundersTaylor1996} applied a similar mean colour--distance 
relation to the construction of subsamples from QDOT because they
thought such a radial gradient of temperature (colour) reflects the intrinsic
evolution. They found that after this correction, the 
clustering difference between ``warm'' and ``cool'' subsamples disappears. Now it is
clear that their procedure of correction actually makes the subsamples
quite contaminated: the nearby galaxies of high $S_{100}/S_{60}$ are likely to be
classified as ``warm'', while those remote galaxies of low colour are marked as ``cool''. 

\subsection{Colour subsamples}
\begin{figure}
\resizebox{\hsize}{!}{\includegraphics{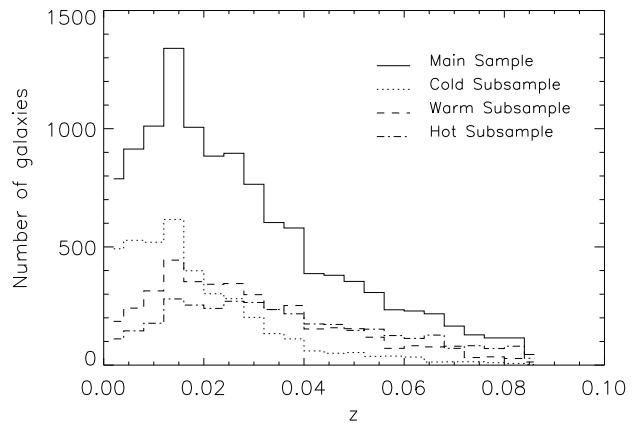}}
\caption{Redshift distributions of galaxies of the three subsamples.}
\label{fig:nz}
\end{figure}

Colour subsamples are generated using the method of \citet{HawkinsEtal2001}. 
``Cold'' galaxies are defined as $S_{100}/S_{60}\ge 2.3$, the
``warm'' subsample consists of those of $1.7\le S_{100}/S_{60}<2.3$, and 
the hot galaxies are the remaining galaxies with $S_{100}/S_{60}< 1.7$. 
Finally, there are 3917, 4010, and 3536 galaxies in the ``cold'', ``warm'' and 
``hot'' subsamples, respectively. Differences in numbers with 
\citet{HawkinsEtal2001} are due to our distance cuts.
The redshift distributions of galaxies of these colour subsamples are 
in Fig.~\ref{fig:nz}.

To estimate the error bars, we also created 20 bootstrap subsamples 
for each colour subsample, as well as the main sample. Scatters at the 
1-$\sigma$ level over the 20 bootstrap subsamples are taken as our errors.

\section{Multifractal analysis}
\subsection{The statistics and selection function}

The multifractal measure in use is constructed from the partition function
\begin{equation}
Z(q,r)=\frac{1}{N}\sum_{i=1}^{N} p_i(r)^{q-1}\propto r^{\tau (q)}\ ,
\label{eq:ci}
\end{equation}
with $p(i)=n_i(r)/N$, where $n_i(r)$ is the count of objects in
the cell of radius $r$, centred upon an object labelled by $i$
(not included in the count), after corrections for the boundary effect and
selection function $\psi(r)$ ($\psi$=1 for volume-limited sample),
\begin{equation}
n_i(r)=\frac{1}{f_i(r)} \sum_{j=1}^{N} \frac{\Psi(|\textbf{r}_j -
\textbf{r}_i| - r)}{\psi(r_j)}\ ,
\label{eq:counts}
\end{equation} 
where
\begin{equation}
\Psi(x)=\left\{\begin{array}{cc} 1, & x \leq 0\\ 0, & x> 0
\end{array}\right. \ ,
\end{equation}
and $f_i(r)$ is the volume fraction of the sphere centred on the
object of radius $r$ within the boundaries of the sample. From 
previous works, we know that on scales larger than about $20\sim 30h^{-1}$Mpc,
the distributions of galaxies are homogeneous \citep{PanColes2000}; the 
comparison here focuses on scales less than $10h^{-1}$Mpc to ensure reliable 
scaling features. Since the boundary correction does not significantly
modify the result for the PSCz sample, especially on small scales \citep{PanColes2002}, 
to reduce the computation burden, the volume correction method is adopted to deal 
with boundaries.

The R\'enyi dimension $D_q$ is derived from the mass exponent $\tau(q)$ by
\begin{equation}
D_q=\tau(q)/(q-1)\ .
\end{equation}
When $q=2$, the partition function $Z$ is the correlation integral and
$D_2$ is the correlation dimension. For each value of $q$, $Z(q)$ and $D_q$ 
give information about the scaling properties of different aspects of the density field.
For high $q$, they are dominated by high-density regions, while for low $q$ and 
$q<0$ the measure is weighted toward low-density regions. Note for $q=1$, the partition
function is defined differently as $Z(q=1,r)=\sum_i \log p(i) \sim r^{D_1}$ 
\citep[cf.][]{MartinezSaar2002}. In this paper, we concentrate on the moments of 
order $q\ge 2$.

\begin{figure}[b]
\resizebox{\hsize}{!}{\includegraphics{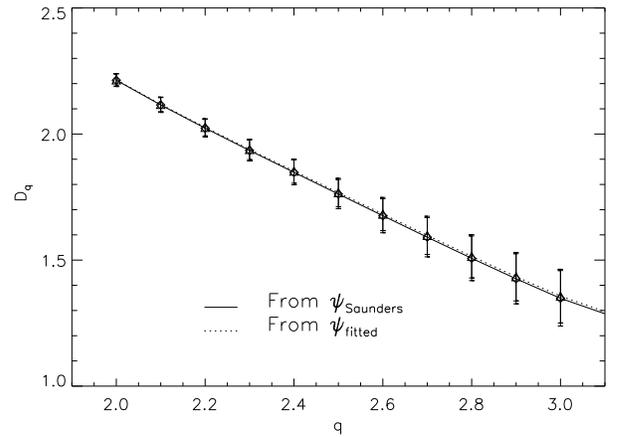}}
\caption{Dimension spectrum of $q\in [2,3]$ of the main sample under 
corrections of different selection functions. One is from
\citet{SaundersEtal2000} and the other is acquired from a
direct numerical fit of $N$-$z$. 
The generalised dimensions are fitted at scales $r<10h^{-1}$Mpc. }
\label{fig:sf}
\end{figure}

From Fig.~\ref{fig:nz} we know that the redshift distributions of the
galaxies of the subsamples are not the same. 
Selection functions of subsamples have to be estimated separately. 
Of course, it is optimal to use a maximum-likelihood method to generate
selection functions for all subsamples. Here we just simply smooth those
curves and then fit the distance-density relation via a nonlinear
least-square fitting to a double power-law function, as in 
\citet{SaundersEtal2000},
\begin{equation}
\psi(r)=\psi_*\left(\frac{r}{r_*}\right)^{1-\alpha}\left[1+
\left(\frac{r}{r_*}\right)^\gamma\right]^{-(\beta/\gamma)} \ ,
\label{eq:selfun}
\end{equation}
in which $\psi_*$, $\alpha$, $r_*$, $\gamma$, and $\beta$ are fitting
parameters. A comparison of
the partition functions of the main sample with the selection function
from \citet{SaundersEtal2000} indicates the accuracy of 
this ``naive'' method and that our simple fits are good enough for the 
statistics (see Fig.~\ref{fig:sf}). 
So, for all colour subsamples, we adopt the nonlinear least-square fitted 
selection functions.

\subsection{Results of multifractal analysis}

\begin{figure*}
\resizebox{\hsize}{!}{\includegraphics{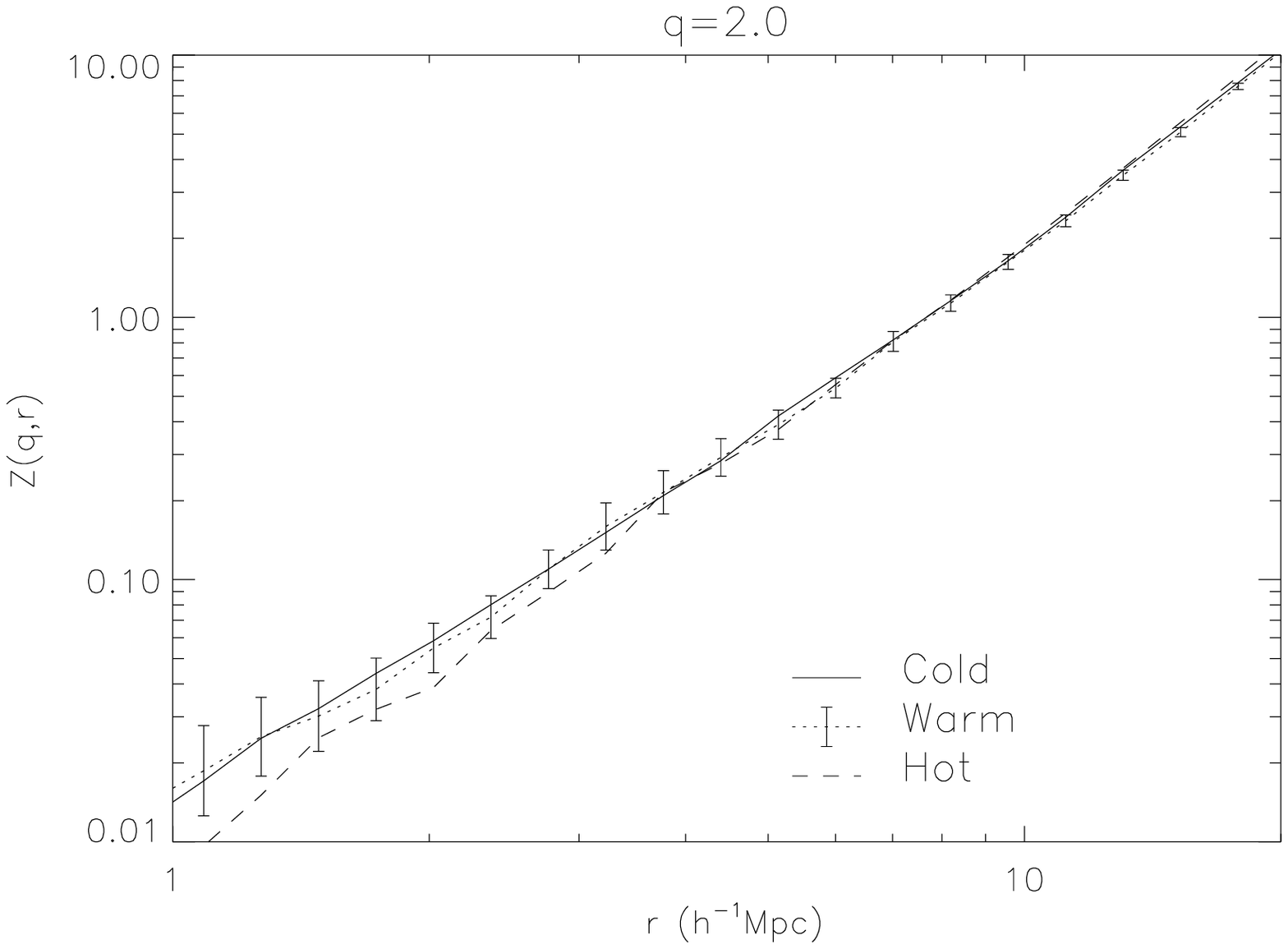}\includegraphics{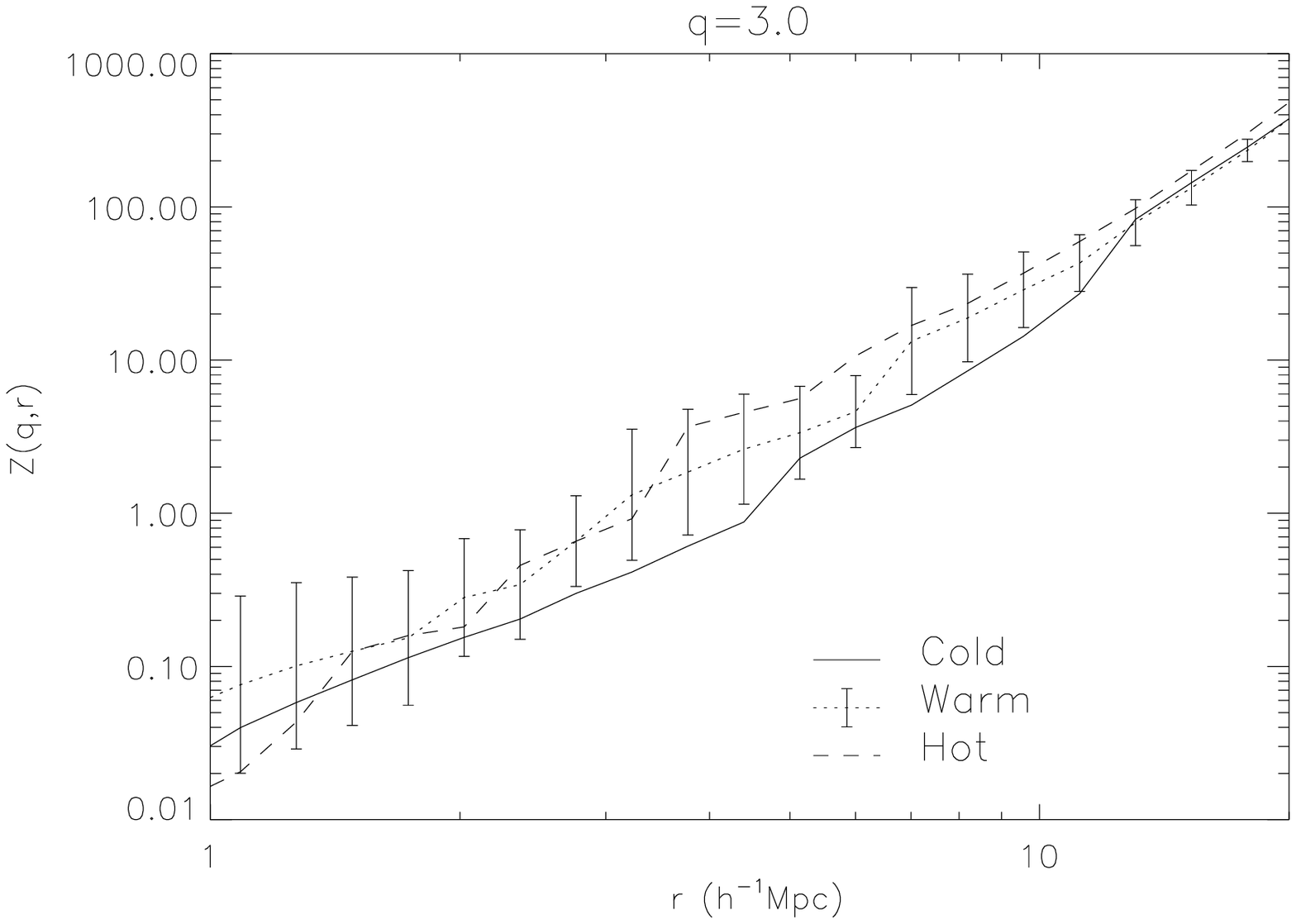}}\\
\resizebox{\hsize}{!}{\includegraphics{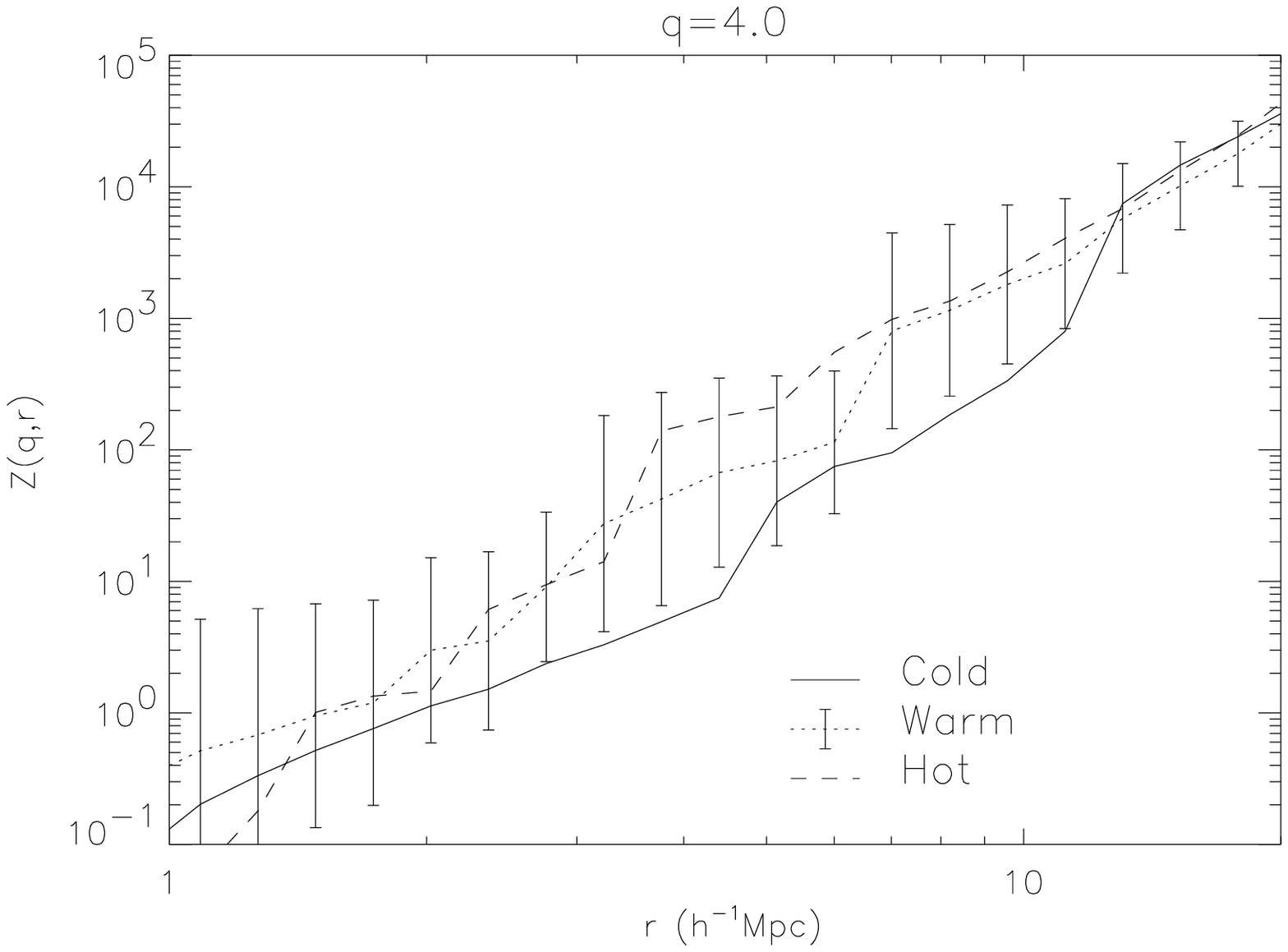}\includegraphics{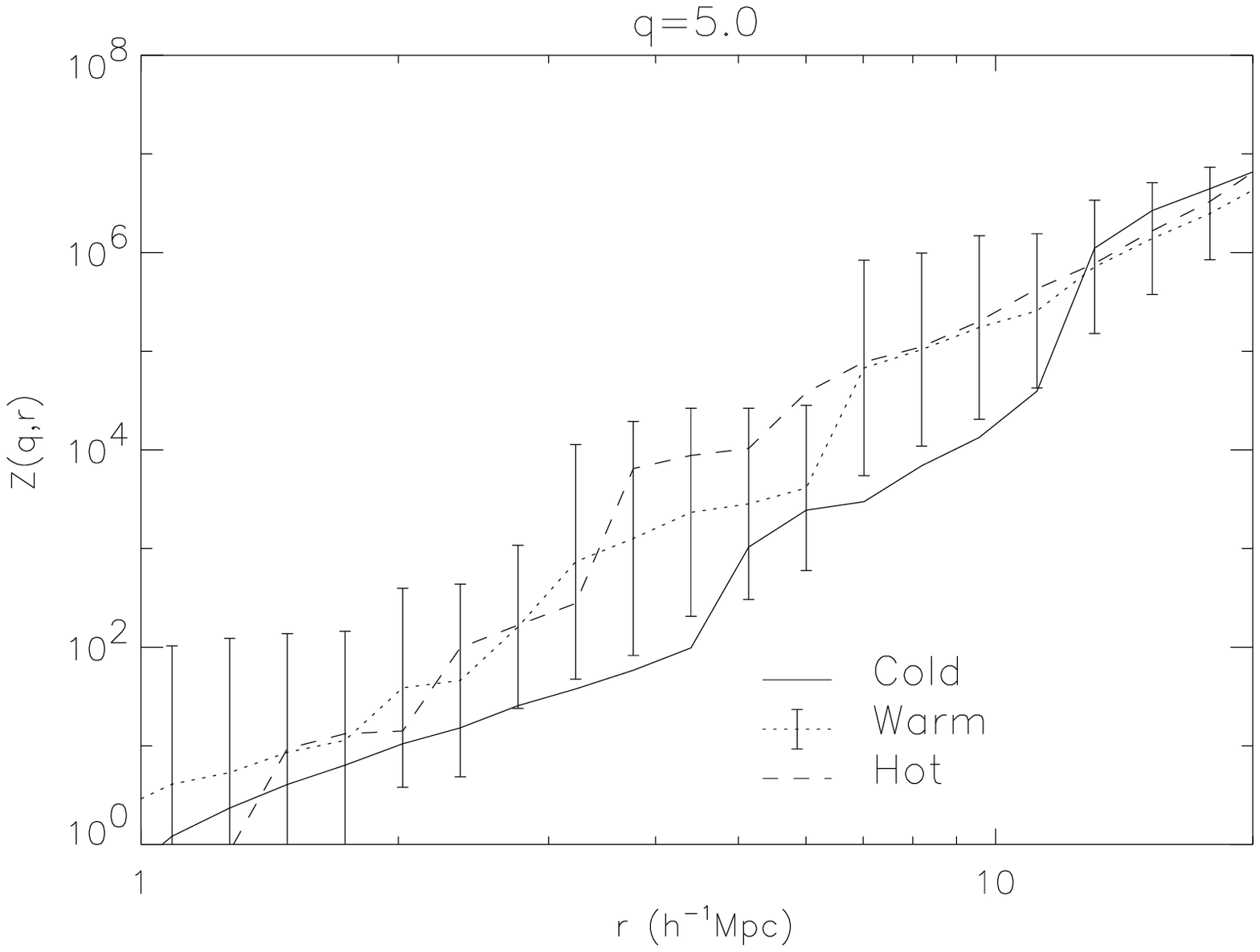}}
\caption{$Z(q, r)$ of q=2, 3, 4, and 5 for colour subsamples.
Only error bars of the warm subsample are plotted; the other two subsamples have
approximately the same size error bars. 
}
\label{fig:ci}
\end{figure*}

\begin{figure}
\resizebox{\hsize}{!}{\includegraphics{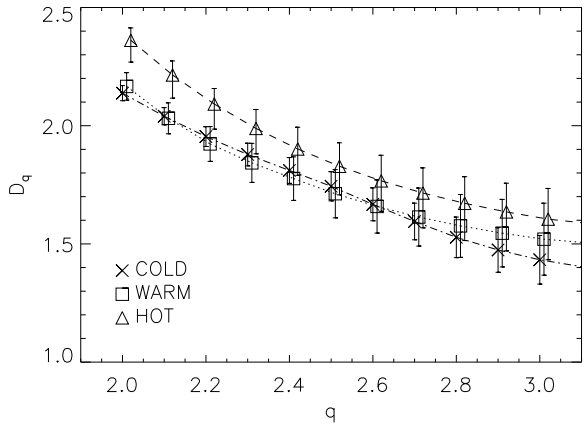}}
\caption{Generalised dimensions $D_q$ of $q\in [2,3]$ in $r\in (1.5, 10h^{-1}{\rm Mpc})$ 
for colour subsamples; points of the ``warm'' and the ``hot'' subsamples are slightly shifted to 
the right for better presentation.}
\label{fig:dq}
\end{figure}

Measurements of $Z(q, r)$ at different orders of our colour subsamples are
displayed in Fig.~\ref{fig:ci}; error bars are estimated from 20 bootstrap subsamples 
for each colour subsample.
It is obvious that on large scales the distributions of all subsamples become 
homogeneous, which means the generalised dimensions are all close to $3$. 
The scales of interests are therefore below $\sim 20 h^{-1}$Mpc, a level at which
the distribution is far from homogeneity. On the other hand, to avoid severe
discreteness effects, measurements on scales of less than $1h^{-1}$Mpc are also
cut off.
In the regime of $\sim 1-10 h^{-1}$Mpc, local
dimensions are approximately constant with small fluctuation,
and the generalised dimensions $D_q$ are calculated
by fitting the partition function in this scale range.
Note that $Z(q, r)$, on small scales of 
all the colour subsamples, appears irregular with large error bars 
when $q>3$, so one should be careful to use the 
dimensionality obtained by a regression of $\log Z(q>3) - \log r$.

Comparisons of the $D_q$ of the subsamples are displayed in Fig.~\ref{fig:dq}. 
The most important dimension is the correlation 
dimension $D_2$. For the ``hot'', ``warm'', and ``cold'' subsamples, $D_2$ is 
$2.34\pm 0.07$, $2.16\pm 0.06$, and $2.14\pm 0.03$, respectively. 
Therefore, cold galaxies are distributed with a smaller fractal
dimension than hot galaxies, which tells us that cold galaxies are more
strongly clustered. 

The dimensionality differences among the three subsamples become smaller for 
higher order moments. High order moments are dominated by those cells from regions with large
number of galaxies, while contributions from cells in sparse regions is effectively suppressed.
So, it is possible that the richest regions of all the subsample 
obey nearly the same scaling law. Unfortunately, the method we used here
is unable to estimate the other half of the dimension spectrum of $q<2$, 
otherwise, it is probable that  we would find bigger differences. Of course, due to 
large uncertainties, one should be conservative about this observation.

Although the partition functions of all three subsamples seem to
asymptotically agree with each other on large scales within error bars, their
selection functions and corresponding normalisation factors are 
different, so we cannot compare the amplitudes of the partition functions
directly.

\section{Conditional multifractal measure}
\subsection{Definition}
It is a tradition to split a main sample into several subsamples
to explore differences in their spatial distributions. This treatment
wastes information provided by the configuration of relative positions
among galaxies of different subsamples. 
For many years the only tool to overcome this
was the cross-correlation function. The mark correlation functions, an 
elegant measure recently introduced
into statistical cosmology are ideal for quantifying the morphology 
dependence of clustering 
\citep{BeisbartKerscher2000, SzapudiEtal2000, Sheth2005}. Using
the mark correlation functions it has already been found that there is no
luminosity dependence of clustering in the PSCz catalogue \citep{SzapudiEtal2000}.

The environment of galaxies of a particular type is a concept closely related 
to cross-correlation. The partition sum of Eq.~\ref{eq:ci} 
measures moment of the counts of neighbours; it is naturally a good tool 
to study the environment, if we include galaxies of all types as neighbours
instead of only galaxies of the same type as the centre galaxy. 
The new measure is given a name of ``conditional multifractal'' 
\citep{RiediScheuring1997}. In the following, we give its formal
definition.

Let the main sample of $N$ points be $\ms=\cup_{m=1}^M S_m$, of which 
$S_{m=1,...,M}$ is a subsample in which there are $N_m$ objects marked by 
their position vectors ${\bf r}_{m_i}$ and
$\{ {\bf r}_j|j=1,..., N \}=\cup_{m=1}^M\{ {\bf r}_{m_i}|i=1,...,N_m \}$.
The conditional partition function of $S_m$ is thus 
\begin{equation}
Z_c^{(m)}(q,r)=\frac{1}{N_m}\sum_{i=1}^{N_m} \left[\frac{\sum_{j=1,j \neq m_i}^{N}
\Psi(|{\bf r}_j-{\bf r}_{m_i}|-r)}{N}\right]^{q-1} .
\end{equation}
For our flux limited sample, we need to apply the same 
corrections to the kernel that we did in Eq.~\ref{eq:counts}. 

Formally, by writing the measure of neighbours in the $j^{th}$ cell as 
$\mu_j, j=1,..., N$, the conditional mass exponent $\tau_c$ and the spectrum of 
the generalised conditional dimension $D_c(q)$ of the subsample $S_m$ is
\begin{equation} 
\tau_c^{(m)}(q)\equiv\frac{d\log Z_c^{(m)}}{d\log r}=\frac{d\log \sum_{i=1}^{N_m}\mu_{m_i}^{q-1}}
{d\log r}\ , \ 
D_c^{(m)}(q)\equiv\frac{\tau_c^{(m)}}{q-1} \ .
\end{equation}
When $q=2$, $Z_c(q=2, r)$ is the conditional correlation integral and $D_c(2)$ is 
the conditional correlation dimension. 

The subset is a collection of specially picked sampling points of the main set, since
\begin{equation}
\sum_{j=1}^{N}\mu_j^{q-1}=\sum_{m=1}^M\sum_{i=1}^{N_m}\mu_{m_i}^{q-1}\ ;
\end{equation}
the relation between $D_q$ and $D_c(q)$ is
\begin{equation}
D_q=\sum_{m=1}^M
\left[ \frac{\sum_{i=1}^{N_m}\mu_{m_i}^{q-1}}{\sum_{j=1}^{N}\mu_j^{q-1}}
\times D_c^{(m)}(q) \right] \ .
\end{equation}
If a subsample $S_m$ is an uniform dilution of $\ms$, 
$\sum_{i=1}^{N_m}\mu_{m_i}^{q-1}/\sum_{j=1}^{N}\mu_j^{q-1}=N_m/N$
and $D_c^{(m)}(q)=D_q$. Any deviation from such uniformity is
embedded in the conditional dimension; the conditional multifractal measure is 
therefore an effective tracer of differences among subsamples.

\subsection{Results of conditional multifractal analysis}

\begin{figure*}
\resizebox{\hsize}{!}{\includegraphics{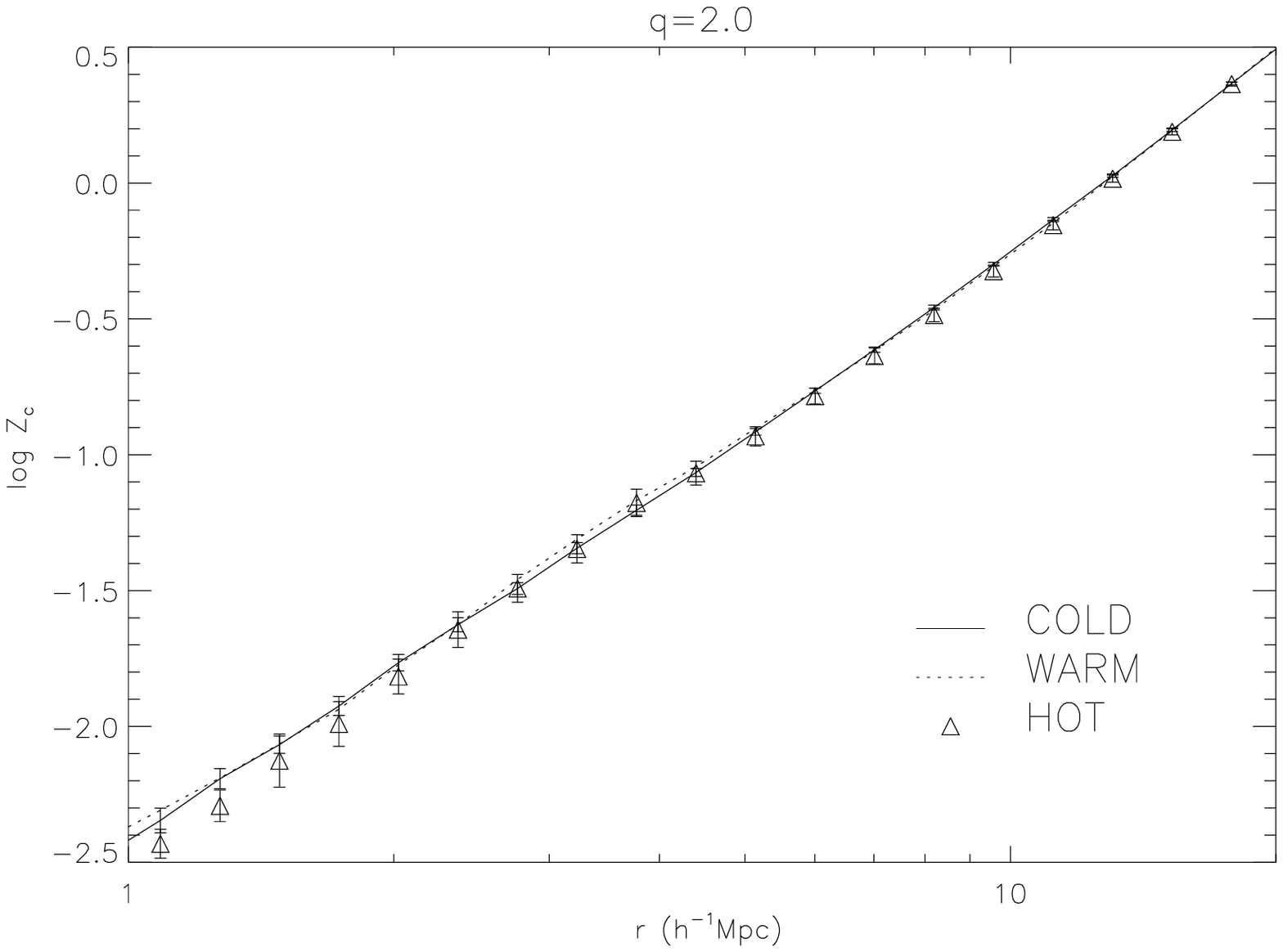}\includegraphics{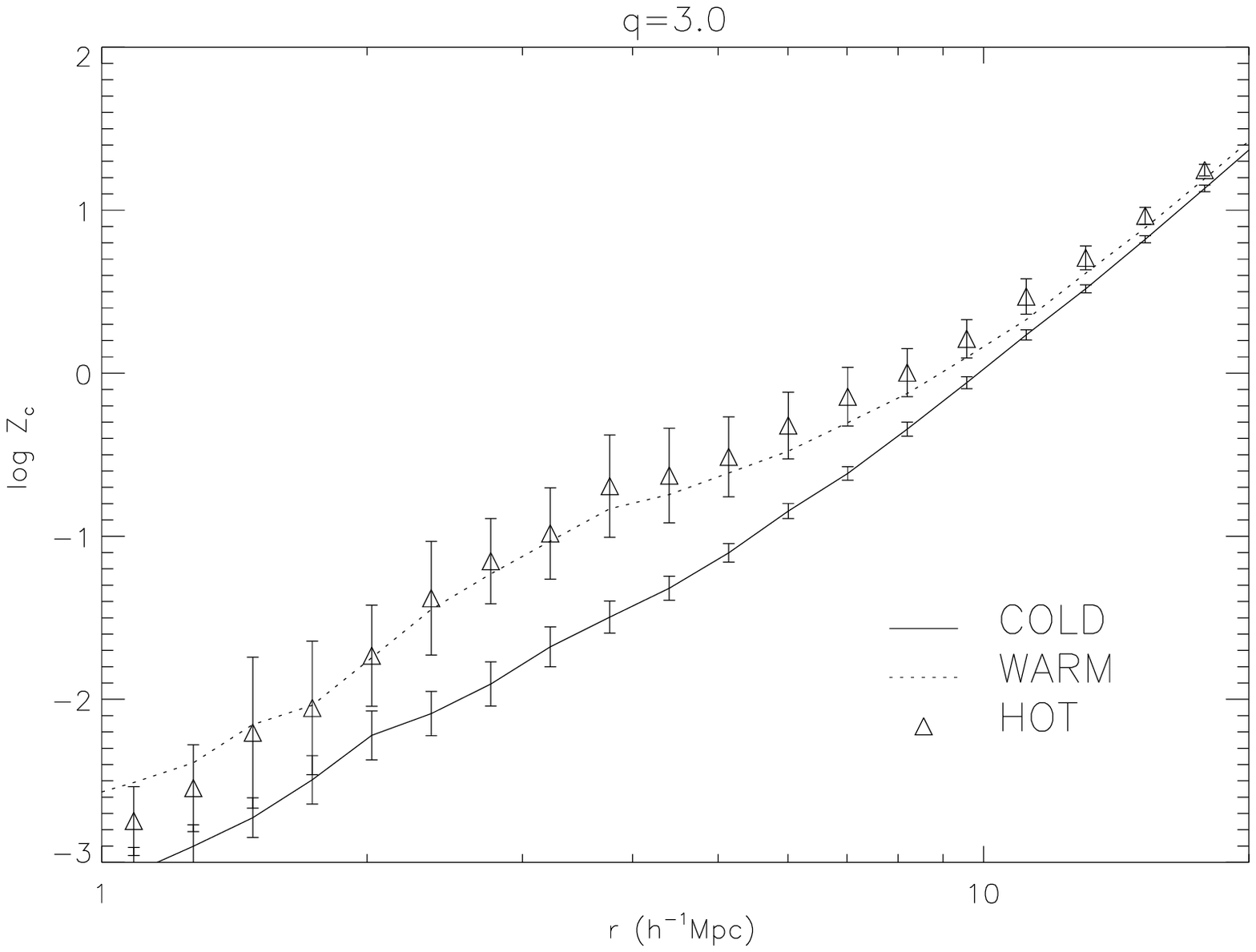}}\\
\resizebox{\hsize}{!}{\includegraphics{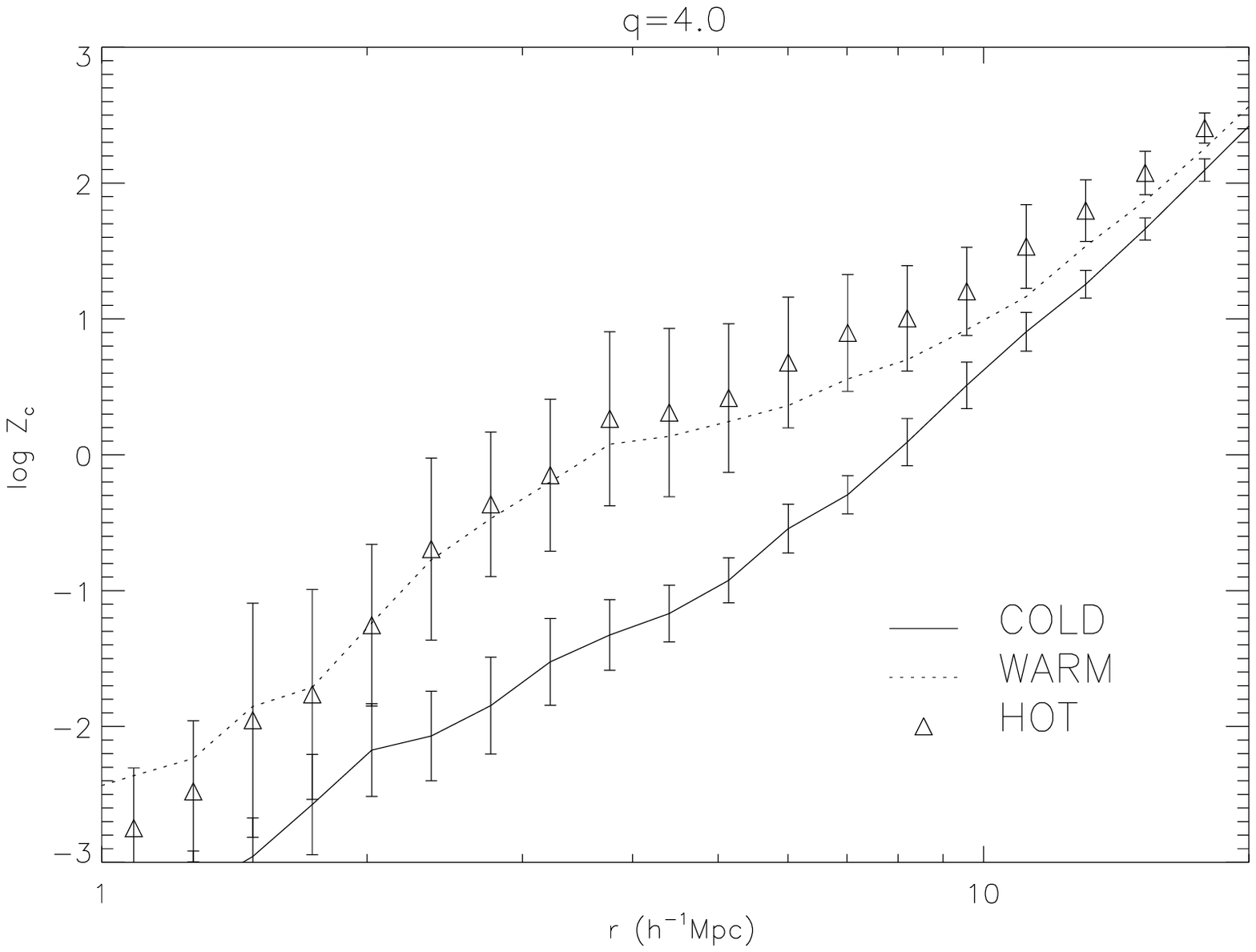}\includegraphics{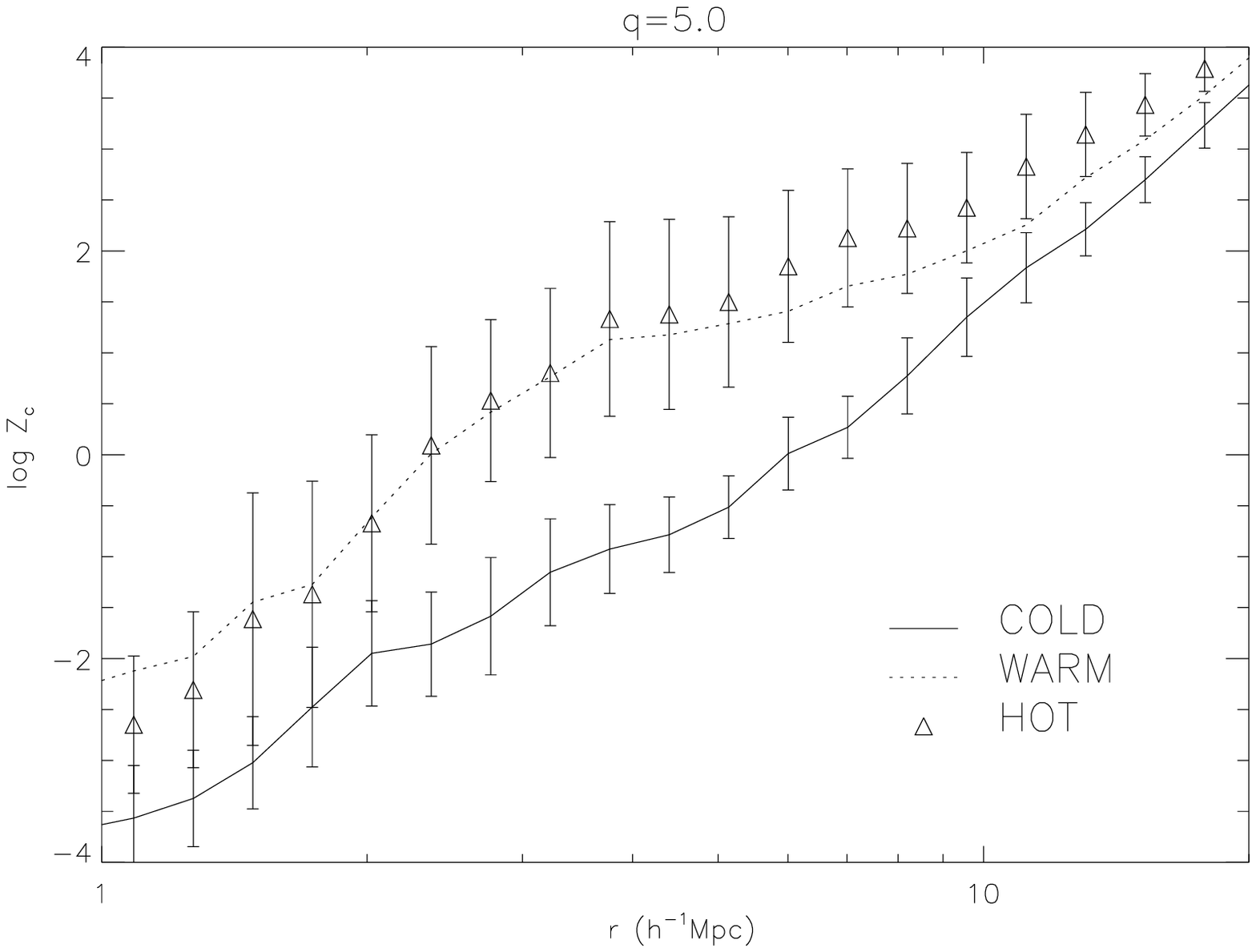}}
\caption{$Z_c(q, r)$ of q=2, 3, 4, and 5 of the PSCz colour subsamples. Error bars 
of the ``warm'' subsample are not plotted; they are very similar in size to 
those of the ``hot'' subsample.}
\label{fig:cci}
\end{figure*}

\begin{figure}
\resizebox{\hsize}{!}{\includegraphics{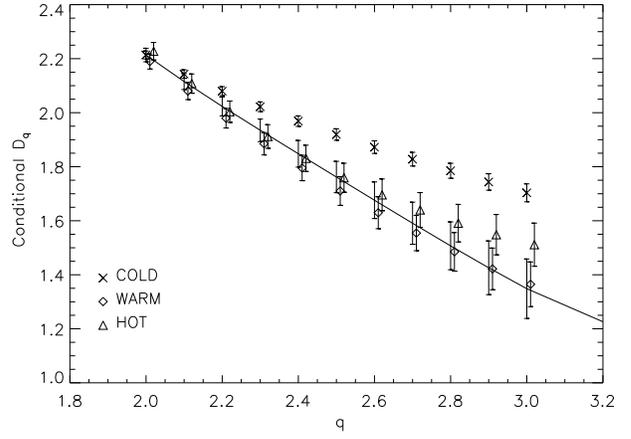}}
\caption{The conditional dimension spectrum within $q\in [2,3]$ and 
$r\in (1,10)h^{-1}$Mpc. Points of the ``warm'' and the ``hot'' subsamples 
are slightly shifted to the right. The solid line with error bars is the 
dimension spectrum of the main sample.}
\label{fig:cdq}
\end{figure}

Conditional partition functions and conditional generalised dimensions 
are calculated for each colour subsample, and
results are displayed in Figs.~\ref{fig:cci} and~\ref{fig:cdq}. Recall that 
neighbours of an object are those in the main sample, so only the selection 
function provided by \citet{SaundersEtal2000} is needed. The error bars are 
from 20 bootstrap subsamples.

Contrary to the ordinary multifractal measure, the conditional
dimensions $D_c(2)$ of the colour subsamples are consistent with each other 
within error bars, while higher orders differ.
We can see that conditional dimension spectrum of the ``hot'' and the ``warm'' 
subsamples are in good agreement with the spectrum of the main sample within 
error bars, which means the two subsamples are basically uniformly diluted 
realisations of the main sample. The ``cold'' subsample has a quite different 
dimension spectrum; its conditional dimensions are larger than those of the 
``hot'' subsample at high orders,
e.g., for q=3, $D_c^{(cold)}=1.70\pm0.04$, and $D_c^{(hot)}=1.50\pm0.08$.
One thing that needs to be addressed is that while the conditional dimensions are 
a measure of the environment of the central object, a larger conditional 
dimension does not mean that the environment is less clustered, it just tells 
us that the class of objects that are measured
are in regions of a special scaling property or ``strangeness'', in fractal
language.

On large scales, the distribution of galaxies is asymptotically
homogeneous, and the $Z_c(q, r)$ of different subsamples are consistent within 
error bars (Fig.~\ref{fig:cci}). So, if the averaged environment of subsamples 
on small scales is different, it will be shown by the amplitudes of $Z_c(q, r)$. 
For q=2, $Z_c$, the mean number of neighbours, of all the colour subsamples agree 
with each other within error bars. When $q>2$, $Z_c^{(cold)}$ is much smaller 
than $Z_c^{(hot)}$. 
Since $\langle \mu\rangle_{cold}=\langle \mu\rangle_{hot}$, the variance in the 
number of neighbours around cold galaxies 
$\langle (\mu-\langle \mu \rangle)^2\rangle_{cold}=
Z_c^{(cold)}(q=3)-Z_c^{(cold)2}(q=2)$ is smaller than that around the hot and 
the warm galaxies, hence the skewness.

\begin{figure}
\resizebox{\hsize}{!}{\includegraphics{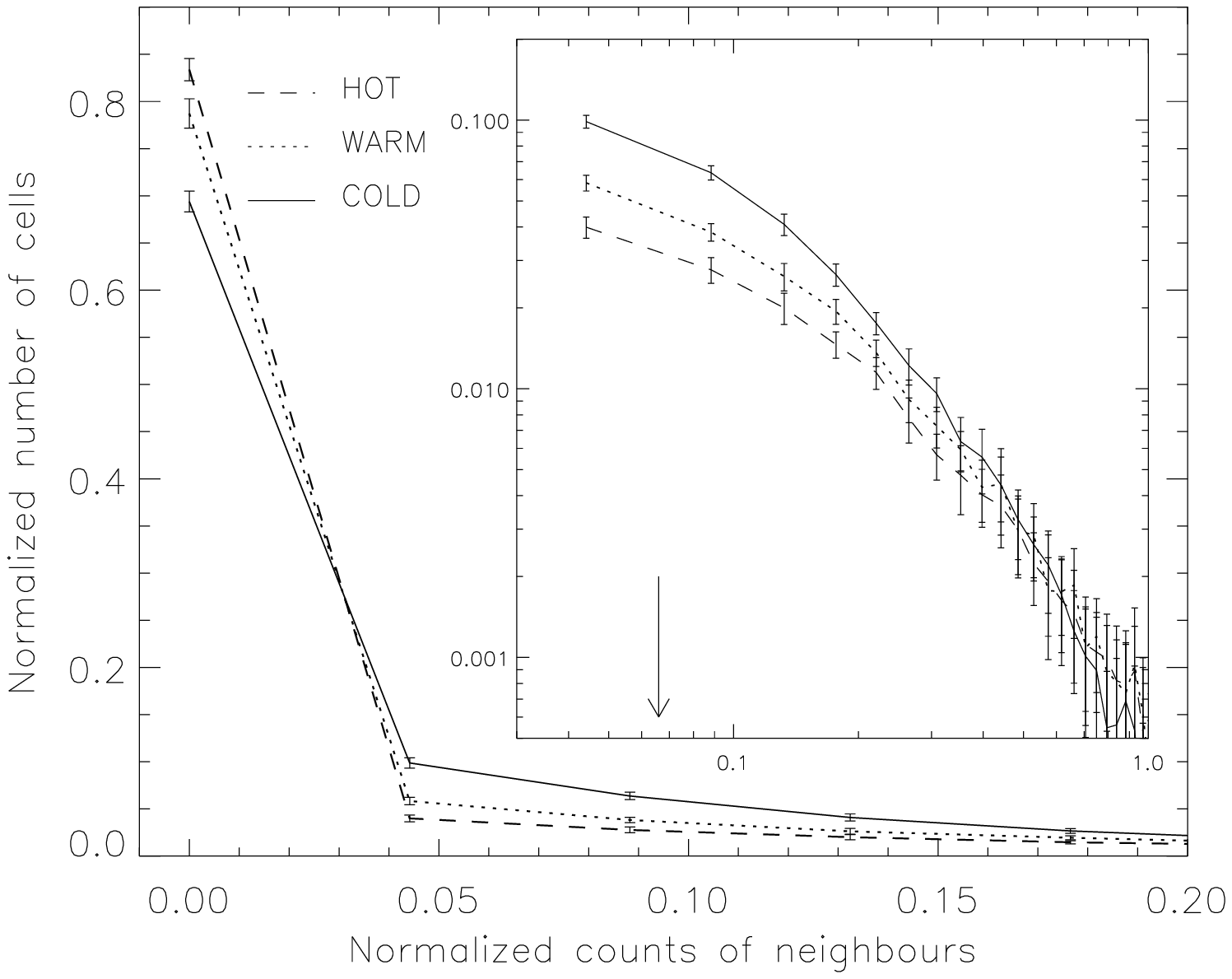}}
\caption{Distribution of the number of neighbours of galaxies in different 
colour subsamples; the inset plot is the log-log version for a better 
demonstration, and the vertical arrow in the 
subplot marks the average normalised neighbour count for a randomly selected galaxy 
belonging to the main sample. The radius of the cell is $r=3.77h^{-1}$Mpc. Counts 
of neighbours are normalised by the number of all galaxies, which is 11463, here. 
The number of cells is counted within a bin of width 0.044, then it is normalised
for each colour subsample by the total number of galaxies of that subsample. 
Error bars are estimated over 20 bootstrap subsamples.}
\label{fig:distri}
\end{figure}

At high $q$, $Z_c(q,r)$ has similar properties to the ordinary $Z(q,r)$ and is also
weighted toward rich cells. It is interesting to see the distribution of richness 
around galaxies. As an example, the distribution function of the number of 
neighbours around galaxies of colour subsamples at scale  $r\sim 4h^{-1}$Mpc 
are shown in Fig.~\ref{fig:distri}. Clearly, hot and warm galaxies have
higher possibilities of being located 
in very sparse regions than cold galaxies, and cold galaxies are more likely 
distributed in rich environments. Cases at other scales less than $10h^{-1}$Mpc 
are similar.

\section{Discussion}
In this work, colour subsamples of PSCz galaxies are analysed with two kinds of 
multifractal measures; all analysis clearly denotes that their spatial distributions are 
different. The ordinary multifractal measure of our three colour subsamples
gives results consistent with the normal morphological segregation scenario of
galaxy clustering. Hot galaxies are clustered less strongly with
a larger correlation dimension $D_2$ than cold galaxies. The difference is
not very strong, and actually, higher-order dimensions fail to detect a
significant difference. 

Details of the spatial distribution of galaxies of different colours
are more clearly displayed by the new statistics --- the conditional 
multifractal that is designed to
measure the environment around galaxies of a particular type. The 
conditional correlation integrals and conditional correlation dimensions are the
same for all colour subsamples within error bars; at high orders $q>2$,
conditional multifractal dimensions detect marginal differences. It is found that 
cold galaxies are in regions with less ``strangeness'' than hot galaxies.

If the model of IRAS galaxies by \citet{RowanCrawford1989} is reasonable, 
the colour defined in the paper is closely associated with the SFR of galaxy.
Effects of environment on star formation activity are very complex under competing
mechanisms. In rich environments, spirals exhibit significant gas deficiencies 
resulting from ram pressure from intracluster mediums, together with possible tidal 
stripping from interactions with other galaxies and cluster potential; in this way, 
star formation rate is effectively suppressed. 
On the other hand, there are external environmental influences that 
can have much stronger effects on boosting the star formation rate. Studies of 
$H\alpha$ and FIR emission of interacting and merging galaxies have shown strong 
excess star formation \citep[see review of][]{Kennicutt1998}. Numerical simulations also
pointed out pair interaction, and merging can produce tidal gas inflows during orbit
decay; the inflow will drastically boost the star formation
\citep[e.g.,][]{BarnesHernquist1996, Tissera2000, TisseraEtal2002}.

As displayed in Fig.~\ref{fig:distri}, in all subsamples, the possibility of finding 
an isolated galaxy is much higher than having galaxies with neighbours, while
the ``hot'' subsample has a larger portion of galaxies with no companion than the ``cold''
subsample. The poor possibility of finding a galaxy with prominent present star formation
activities in a rich environment can be interpreted as follows: in crowded regions, most galaxies 
experienced an active star formation stage on a very short time scale at high redshift,
and then were quenched afterwards. The galaxies that were still there with considerable 
present SFR are likely the small number of survivors. This conclusion is similar to
the analysis of the 2dFGRS and SDSS surveys by \citet{BaloghEtal2004}. 
Actually, observations of Spitzer have shown that the bulk of star formation in massive 
galaxies occurs at early cosmic epochs and is largely complete by $z\sim1.5$, while at $z<1$ 
lower-mass galaxies dominate the overall cosmic mass assembly \citep{Papovich2006}. It 
is highly possible that the IRAS galaxies in rich regions are of small mass, which is
consistent with the discovery of \citet{MartinezEtal2002}. Another aspect is that the field
IRAS galaxies are probably systems of slow evolution, which may be tested by studies on
the SFRs of galaxies with a ``loose'' friend and galaxies without a neighbour over
a period of time.

Nevertheless, in Fig.~\ref{fig:distri}, with the increasing number of neighbours, 
the probability of a hot galaxy living with such richness is smaller than that
of a cold galaxy. This picture shows the downside effects of environment on the SFRs of 
galaxies, which is consistent with recent discoveries that high SFR galaxies are inhibited
in rich environments in the 2dFGRS and SDSS surveys \citep{MartinezEtal2002, DominguezEtal2002, 
BaloghEtal2004}.

In the high richness regime, the distribution of hot and cold galaxies are asymptotically 
in agreement with each other within error bars. There must also be some mechanism 
that effectively enhances the galaxy's star formation rate,
partly in compensation for the general suppression
when a region becomes very crowded, since otherwise the distribution curve of the 
``hot'' subsample should be always lower than the ``cold'' subsample. 
\citet{KrongoldEtal2002} found that interacting IRAS galaxies have lower $S_{100}/S_{60}$ 
than isolated galaxies. It has also been found that in very dense regions,
there is induced star formation in galaxy pairs at a very small separation,
due to interactions, and that a galaxy with a high star formation rate is likely to 
trigger more star formation in its close companion 
\citep[e.g.,][]{CarterEtal2001, TisseraEtal2002, LewisEtal2002, AlonsoEtal2004,AlonsoEtal2006}. 
However, in this regime, our distribution curves have large 
uncertainties as there are not many cells within the bin for counting since IRAS galaxies
avoid rich clusters. We shall be conservative about this claim, which needs confirmation
from future works with more complete and denser samples than the PSCz.

\begin{acknowledgement}
The author thanks the anonymous referee for very helpful comments, and appreciates useful 
suggestions from Peter Coles. The author also acknowledges the hospitality and kindness 
of the Astronomy Department of Peking University and the Beijing Astronomical Planetarium.
\end{acknowledgement}


\begin{thebibliography}{48}
\expandafter\ifx\csname natexlab\endcsname\relax\def\natexlab#1{#1}\fi

\bibitem[{{Balogh} {et~al.}(2004){Balogh}, {Eke}, {Miller}, {Lewis}, {Bower},
  {et~al.}}]{BaloghEtal2004}
{Balogh}, M., {Eke}, V., {Miller}, C., {et~al.} 2004, \mnras, 348, 1355

\bibitem[{{Balogh} {et~al.}(1998){Balogh}, {Schade}, {Morris}, {Yee},
  {Carlberg}, \& {Ellingson}}]{BaloghEtal1998}
{Balogh}, M.~L., {Schade}, D., {Morris}, S.~L., {et~al.} 1998, \apjl, 504, L75

\bibitem[{{Bardeen} {et~al.}(1986){Bardeen}, {Bond}, {Kaiser}, \&
  {Szalay}}]{BardeenEtal1986}
{Bardeen}, J.~M., {Bond}, J.~R., {Kaiser}, N., \& {Szalay}, A.~S. 1986, \apj,
  304, 15

\bibitem[{{Barnes} \& {Hernquist}(1996)}]{BarnesHernquist1996}
{Barnes}, J.~E. \& {Hernquist}, L. 1996, \apj, 471, 115

\bibitem[{{Beisbart} \& {Kerscher}(2000)}]{BeisbartKerscher2000}
{Beisbart}, C. \& {Kerscher}, M. 2000, \apj, 545, 6

\bibitem[{{Best} {et~al.}(1996){Best}, {Charlton}, \&
  {Mayer-Kress}}]{BestEtal1996}
{Best}, J.~S., {Charlton}, J.~C., \& {Mayer-Kress}, G. 1996, \apj, 456, 55

\bibitem[{{Blanton} {et~al.}(1999){Blanton}, {Cen}, {Ostriker}, \&
  {Strauss}}]{BlantonEtal1999}
{Blanton}, M., {Cen}, R., {Ostriker}, J.~P., \& {Strauss}, M.~A. 1999, \apj,
  522, 590

\bibitem[{{Caon} \& {Einasto}(1995)}]{CaonEinasto1995}
{Caon}, N. \& {Einasto}, M. 1995, \mnras, 273, 913

\bibitem[{{Carter} {et~al.}(2001){Carter}, {Fabricant}, {Geller}, {Kurtz}, \&
  {McLean}}]{CarterEtal2001}
{Carter}, B.~J., {Fabricant}, D.~G., {Geller}, M.~J., {Kurtz}, M.~J., \&
  {McLean}, B. 2001, \apj, 559, 606

\bibitem[{{Dekel} \& {Lahav}(1999)}]{DekelLahav1999}
{Dekel}, A. \& {Lahav}, O. 1999, \apj, 520, 24

\bibitem[{{Dom{\'{\i}}nguez} {et~al.}(2002){Dom{\'{\i}}nguez}, {Zandivarez},
  {Mart{\'{\i}}nez}, {Merch{\'a}n}, {Muriel}, \& {Lambas}}]{DominguezEtal2002}
{Dom{\'{\i}}nguez}, M.~J., {Zandivarez}, A.~A., {Mart{\'{\i}}nez}, H.~J.,
  {et~al.} 2002, \mnras, 335, 825

\bibitem[{{Dom\'inguez-Tenreiro} {et~al.}(1994){Dom\'inguez-Tenreiro},
  {Campos}, {Gomez-Flechoso}, \& {Yepes}}]{DominguezEtal1994}
{Dom\'inguez-Tenreiro}, R., {Campos}, A., {Gomez-Flechoso}, M.~A., \& {Yepes},
  G. 1994, \apjl, 424, L73

\bibitem[{{Dom\'inguez-Tenreiro} \& {Mart\'inez}(1989)}]{DominguezMartinez1989}
{Dom\'inguez-Tenreiro}, R. \& {Mart\'inez}, V.~J. 1989, \apjl, 339, L9

\bibitem[{{Dressler}(1980)}]{Dressler1980}
{Dressler}, A. 1980, \apj, 236, 351

\bibitem[{{Dressler} {et~al.}(1997){Dressler}, {Oemler}, {Couch}, {Smail},
  {Ellis}, {Barger}, {Butcher}, {Poggianti}, \& {Sharples}}]{DresslerEtal1997}
{Dressler}, A., {Oemler}, A.~J., {Couch}, W.~J., {et~al.} 1997, \apj, 490, 577

\bibitem[{{Guzzo} {et~al.}(2000)}]{GuzzoEtal2000}
{Guzzo}, L. {et~al.} 2000, \aap, 355, 1

\bibitem[{{Hawkins} {et~al.}(2001){Hawkins}, {Maddox}, {Branchini}, \&
  {Saunders}}]{HawkinsEtal2001}
{Hawkins}, E., {Maddox}, S., {Branchini}, E., \& {Saunders}, W. 2001, \mnras,
  325, 589

\bibitem[{{Hermit} {et~al.}(1996){Hermit}, {Santiago}, {Lahav}, {Strauss},
  {Davis}, {Dressler}, \& {Huchra}}]{HermitEtal1996}
{Hermit}, S., {Santiago}, B.~X., {Lahav}, O., {et~al.} 1996, \mnras, 283, 709

\bibitem[{{Kaiser}(1984)}]{Kaiser1984}
{Kaiser}, N. 1984, \apjl, 284, L9

\bibitem[{{Kennicutt}(1998)}]{Kennicutt1998}
{Kennicutt}, R.~C. 1998, \araa, 36, 189

\bibitem[{{Krongold} {et~al.}(2002){Krongold}, {Dultzin-Hacyan}, \&
  {Marziani}}]{KrongoldEtal2002}
{Krongold}, Y., {Dultzin-Hacyan}, D., \& {Marziani}, P. 2002, \apj, 572, 169

\bibitem[{{Lewis} {et~al.}(2002){Lewis}, {Balogh}, {De Propris}, {Couch},
  {Bower}, {et~al.}}]{LewisEtal2002}
{Lewis}, I., {Balogh}, M., {De Propris}, R., {et~al.} 2002, \mnras, 334, 673

\bibitem[{{Lin} {et~al.}(1996){Lin}, {Kirshner}, {Shectman}, {Landy}, {Oemler},
  {Tucker}, \& {Schechter}}]{LinEtal1996}
{Lin}, H., {Kirshner}, R.~P., {Shectman}, S.~A., {et~al.} 1996, \apj, 471, 617

\bibitem[{{Madgwick} {et~al.}(2003)}]{MadgwickEtal2003}
{Madgwick}, D.~S. {et~al.} 2003, \mnras, 344, 847

\bibitem[{{Mann} {et~al.}(1996){Mann}, {Saunders}, \&
  {Taylor}}]{MannSaundersTaylor1996}
{Mann}, R.~G., {Saunders}, W., \& {Taylor}, A.~N. 1996, \mnras, 279, 636

\bibitem[{{Mart{\'{\i}}nez} {et~al.}(2002){Mart{\'{\i}}nez}, {Zandivarez},
  {Dom{\'{\i}}nguez}, {Merch{\'a}n}, \& {Lambas}}]{MartinezEtal2002}
{Mart{\'{\i}}nez}, H.~J., {Zandivarez}, A., {Dom{\'{\i}}nguez}, M.,
  {Merch{\'a}n}, M.~E., \& {Lambas}, D.~G. 2002, \mnras, 333, L31

\bibitem[{{Mart{\'{\i}}nez} \& {Saar}(2002)}]{MartinezSaar2002}
{Mart{\'{\i}}nez}, V.~J. \& {Saar}, E. 2002, {Statistics of the Galaxy
  Distribution} (Boca Raton: Chapman \& Hall/CRC press)

\bibitem[{{Mo} \& {White}(1996)}]{MoWhite1996}
{Mo}, H.~J. \& {White}, S.~D.~M. 1996, \mnras, 282, 347

\bibitem[{{Norberg} {et~al.}(2002){Norberg}, {Baugh}, {Hawkins}, {Maddox},
  {Madgwick}, {et~al.}}]{NorbergEtal2002a}
{Norberg}, P., {Baugh}, C.~M., {Hawkins}, E., {et~al.} 2002, \mnras, 332, 827

\bibitem[{{Norberg} {et~al.}(2001){Norberg}, {Baugh}, {Hawkins}, {Maddox},
  {et~al.}}]{NorbergEtal2001}
{Norberg}, P., {Baugh}, C.~M., {Hawkins}, E., {Maddox}, S., {et~al.} 2001,
  \mnras, 328, 64

\bibitem[{{Pan} \& {Coles}(2000)}]{PanColes2000}
{Pan}, J. \& {Coles}, P. 2000, \mnras, 318, L51

\bibitem[{{Pan} \& {Coles}(2002)}]{PanColes2002}
---. 2002, \mnras, 330, 719

\bibitem[{{Papovich}(2006)}]{Papovich2006}
{Papovich}, C. 2006, New Astronomy Review, 50, 134

\bibitem[{{Richards} \& {Scheuring}(1997)}]{RiediScheuring1997}
{Richards}, R.~H. \& {Scheuring}, I. 1997, Fractals, 5, 153

\bibitem[{{Rowan-Robinson} \& {Crawford}(1989)}]{RowanCrawford1989}
{Rowan-Robinson}, M. \& {Crawford}, J. 1989, \mnras, 238, 523

\bibitem[{{Saunders} {et~al.}(1990){Saunders}, {Rowan-Robinson}, {Lawrence},
  {Efstathiou}, {Kaiser}, {Ellis}, \& {Frenk}}]{SaundersEtal1990}
{Saunders}, W., {Rowan-Robinson}, M., {Lawrence}, A., {et~al.} 1990, \mnras,
  242, 318

\bibitem[{{Saunders} {et~al.}(2000){Saunders}, {Sutherland}, {Maddox},
  {Keeble}, {Oliver}, {et~al.}}]{SaundersEtal2000}
{Saunders}, W., {Sutherland}, W.~J., {Maddox}, S.~J., {et~al.} 2000, \mnras,
  317, 55

\bibitem[{{Serjeant} {et~al.}(2004){Serjeant}, {Carrami{\~n}ana},
  {Gonz{\'a}les-Solares}, {H{\'e}raudeau}, {M{\'u}jica},
  {et~al.}}]{SerjeantEtal2004}
{Serjeant}, S., {Carrami{\~n}ana}, A., {Gonz{\'a}les-Solares}, E., {et~al.}
  2004, \mnras, 355, 813

\bibitem[{{Shepherd} {et~al.}(2001){Shepherd}, {Carlberg}, {Yee}, {Morris},
  {Lin}, {Sawicki}, {Hall}, \& {Patton}}]{ShepherdEtal2001}
{Shepherd}, C.~W., {Carlberg}, R.~G., {Yee}, H.~K.~C., {et~al.} 2001, \apj,
  560, 72

\bibitem[{{Sheth}(2005)}]{Sheth2005}
{Sheth}, R.~K. 2005, \mnras, 364, 796

\bibitem[{{Sol Alonso} {et~al.}(2006){Sol Alonso}, {Lambas}, {Tissera}, \&
  {Coldwell}}]{AlonsoEtal2006}
{Sol Alonso}, M., {Lambas}, D.~G., {Tissera}, P., \& {Coldwell}, G. 2006,
  \mnras, in press, astro-ph/0511362

\bibitem[{{Sol Alonso} {et~al.}(2004){Sol Alonso}, {Tissera}, {Coldwell}, \&
  {Lambas}}]{AlonsoEtal2004}
{Sol Alonso}, M., {Tissera}, P.~B., {Coldwell}, G., \& {Lambas}, D.~G. 2004,
  \mnras, 352, 1081

\bibitem[{{Szapudi} {et~al.}(2000){Szapudi}, {Branchini}, {Frenk}, {Maddox}, \&
  {Saunders}}]{SzapudiEtal2000}
{Szapudi}, I., {Branchini}, E., {Frenk}, C.~S., {Maddox}, S., \& {Saunders}, W.
  2000, \mnras, 318, L45

\bibitem[{{Tissera}(2000)}]{Tissera2000}
{Tissera}, P.~B. 2000, \apj, 534, 636

\bibitem[{{Tissera} {et~al.}(2002){Tissera}, {Dom{\'{\i}}nguez-Tenreiro},
  {Scannapieco}, \& {S{\'a}iz}}]{TisseraEtal2002}
{Tissera}, P.~B., {Dom{\'{\i}}nguez-Tenreiro}, R., {Scannapieco}, C., \&
  {S{\'a}iz}, A. 2002, \mnras, 333, 327

\bibitem[{{Wen} {et~al.}(1989){Wen}, {Deng}, {Liu}, \& {Xia}}]{WenDengXia1989}
{Wen}, Z., {Deng}, Z.-G., {Liu}, Y.-Z., \& {Xia}, X.-Y. 1989, \aap, 219, 1

\bibitem[{{Young} {et~al.}(1996){Young}, {Allen}, {Kenney}, {Lesser}, \&
  {Rownd}}]{YoungEtal1996}
{Young}, J.~S., {Allen}, L., {Kenney}, J.~D.~P., {Lesser}, A., \& {Rownd}, B.
  1996, \aj, 112, 1903

\bibitem[{{Zehavi} {et~al.}(2005)}]{ZehaviEtal2005}
{Zehavi}, I. {et~al.} 2005, \apj, 630, 1

\end{thebibliography}

\end{document}